\documentclass[12pt,a4paper]{article}
\usepackage{setspace}
\usepackage{subfig}
\usepackage{authblk}
\usepackage{amssymb}
\usepackage{upquote}

\usepackage{babel}
\usepackage{beramono}
\usepackage[T1]{fontenc}
\usepackage[latin9]{inputenc}
\usepackage{color}
\usepackage{rotating, graphicx}

\definecolor{identifiercolor}{rgb}{.4,.6,.56}
\definecolor{stringcolor}{gray}{0.5}
\definecolor{inactivecolor}{rgb}{0.15,0.15,0.5}

\usepackage{listings}
\usepackage{listings}
\usepackage{caption}
\usepackage{amsmath, epsfig,graphicx}
\usepackage{color}
\usepackage{enumerate}

\usepackage{titlesec}
\titleformat{\section}{\normalsize\bfseries}{\thesection}{1em}{}

\def\Title#1{\begin{center} {\Large \bf #1 } \end{center}}
\def\Abstracti#1{\begin{center} {\small \bf #1 } \end{center}}
\def\Author#1{\begin{center}{ \sc #1} \end{center}}
\def\Address#1{\begin{center}{ \it #1} \end{center}}
\newenvironment{Abstract}{\begin{quotation}  }{\end{quotation}}

\begin{document}

\begin{titlepage}
	\let\thefootnote\relax\footnote{* Mansouri@quantgates.co.uk}
	
	\Title{}
	\Title{}
	\Title{A Mini-Introduction To Superfield Decompositions With Branching Rules}
	\vfil
	\Author{ Behzad Mansouri*}
	\vfil
	\Address{QuantGates Ltd, 
		London, EC2A 4NE, UK}
	\vfil
	\Abstracti {Abstract}
	\begin{Abstract}
This paper provides a short introduction to scalar, bosonic, and fermionic superfield component expansion based on the branching rules of irreducible representations in one Lie algebra (in our case, $\mathfrak{su}(32)$, and also $\mathfrak{su}(16)$) into one of its Lie subalgebras ($\mathfrak{so}(11)$, $\mathfrak{so}(10)$). This systematic method paves the way for expansion of bosonic and fermionic superfields, in order to search for possible off-shell supergravity supermultiplets. Furthermore, we implement such a decomposition method in Mathematica in its simplest form, which can be used for superfield component decompositions.
	\end{Abstract}
	\vfill
	\text{Keywords:}  supersymmetry, superfields, off-shell, branching rules, Mathematica
	\vfill

\end{titlepage}

\section{Introduction}
The superfield formalism introduced by A. Salam and J. Strathdee in \cite{Salam:1974yz} allows one to construct supersymmetric models in which supersymmetry becomes inherently manifest. Although supersymmetry manifestation can be achieved with the superfield formalism, full decomposition of unconstrained superfield in higher dimensions via $\theta$-expansions can not be achieved efficiently and until recently, the component-level full structures of superfields in high dimensions had not been observed.
In \cite{JamesGates:2019bwu,JamesGates:2020hrl}, S.J. Gates, Jr, Y. Hu and S.-N. Mak, for the first time utilized systematic branching rules for full expansion of component fields of off-shell superfields in D=10 and D=11. A full scan prescribed for a scalar superfield in 11D, $\mathcal{N}=1$ revealed 1,494 bosonic fields and 1,186 fermionic fields present in the scalar superfield.  
This efficient decomposition method paved the way for studying the expansions of some bosonic and fermionic superfields, in order to find possible off-shell supergravity supermultiplets.

In this paper, we first review the complications arising in the $\theta$-expansion, then provide branching rules implementation in Mathematica, which one can use to derive superfield component decompositions (especially in higher dimensions), as well as to verify the results presented in \cite{JamesGates:2019bwu,JamesGates:2020hrl}.

\section{Complications of $\theta$-Expansion }
In \cite{JamesGates:2020hrl}, the redundancy and inefficiency of the general rules of $\theta$-expansion have been fully discussed.
Here, we quickly review the problem of redundancy, which makes the whole procedure inefficient; particularly in higher dimensions. 
Let us consider the few first terms of a real scalar superfield $S({x^{\underline k }},{\theta ^\alpha })$ in its $\theta$-expansion:

\begin{equation}
	\begin{array}{l}

S({x^{\underline k }},{\theta ^\alpha }) = {s^{(0)}}({x^{\underline k }}) + {\theta ^\alpha }{s_\alpha }^{(1)}({x^{\underline k }})\\
 + {C_{\alpha \beta }}{\theta ^\alpha }{\theta ^\beta }{s^{(2)}}({x^{\underline k }}) + {({\gamma ^{\underline {abc} }})_{\alpha \beta }}{\theta ^\alpha }{\theta ^\beta }{s_{\underline {abc} }}^{(2)}({x^{\underline k }}) + {({\gamma ^{\underline {abcd} }})_{\alpha \beta }}{\theta ^\alpha }{\theta ^\beta }{s_{\underline {abcd} }}^{(2)}({x^{\underline k }})\\
 + {C_{\alpha \beta }}{\theta ^\alpha }{\theta ^\beta }{\theta ^\gamma }{s_\gamma }^{(3)}({x^{\underline k }}) + {({\gamma ^{\underline {abc} }})_{\alpha \beta }}{\theta ^\alpha }{\theta ^\beta }{\theta ^\gamma }{s_{\gamma \underline {abc} }}^{(3)}({x^{\underline k }}) + {({\gamma ^{\underline {abcd} }})_{\alpha \beta }}{\theta ^\alpha }{\theta ^\beta }{\theta ^\gamma }{s_{\gamma \underline {abcd} }}^{(3)}({x^{\underline k }})\\
 + {C_{\alpha \beta }}{\theta ^\alpha }{\theta ^\beta }{C_{\rho \sigma }}{\theta ^\rho }{\theta ^\sigma }{s^{(4)}}({x^{\underline k }}) + {C_{\alpha \beta }}{\theta ^\alpha }{\theta ^\beta }{({\gamma ^{\underline {abc} }})_{\rho \sigma }}{\theta ^\rho }{\theta ^\sigma }{s_{\underline {abc} }}^{(4)}({x^{\underline k }})\\
 + {C_{\alpha \beta }}{\theta ^\alpha }{\theta ^\beta }{({\gamma ^{\underline {abcd} }})_{\rho \sigma }}{\theta ^\rho }{\theta ^\sigma }{s_{\underline {abcd} }}^{(4)}({x^{\underline k }}) + {({\gamma ^{\underline {abc} }})_{\alpha \beta }}{\theta ^\alpha }{\theta ^\beta }{({\gamma ^{\underline {def} }})_{\rho \sigma }}{\theta ^\rho }{\theta ^\sigma }{s_{\underline {abc} }}{_{\underline {def} }^{(4)}}({x^{\underline k }})\\
 + {({\gamma ^{\underline {abc} }})_{\alpha \beta }}{\theta ^\alpha }{\theta ^\beta }{({\gamma ^{\underline {defg} }})_{\rho \sigma }}{\theta ^\rho }{\theta ^\sigma }{s_{\underline {abc} }}{_{\underline {defg} }^{(4)}}({x^{\underline k }}) + {({\gamma ^{\underline {abcd} }})_{\alpha \beta }}{\theta ^\alpha }{\theta ^\beta }{({\gamma ^{\underline {efgh} }})_{\rho \sigma }}{\theta ^\rho }{\theta ^\sigma }{s_{\underline {abcd} }}{_{\underline {efgh} }^{(4)}}({x^{\underline k }})\\
 +  \ldots
	\end{array}
\end{equation}
At level 2, the quadratic basis ${\theta ^\alpha } \wedge {\theta ^\beta }$ are 496-dimensional and all possible quadratic $\theta$-monomials appearing in Equation (1) correspond to $\{330, 165, 1\}$, the bosonic representations, and there is no redundancy at this level.

At level 3, one can conduct the decomposition as \cite{JamesGates:2020hrl}:
\begin{equation}
	\begin{array}{l}

\{ 32\}  \wedge \{ 32\}  \wedge \{ 32\}  = \{ 4,960\}  = \{ 32\}  \oplus \{ 1,408\}  \oplus \{ 3,520\} 

	\end{array}
\end{equation}

For the spinorial representation $\{ 32\}$, one can find three types of cubic $\theta$-monomials with one free spinor index and no free vector index as follows (using the notation of \cite{JamesGates:2020hrl}):

\begin{equation}
\begin{array}{l}
T1 = \gamma _{\delta \varepsilon }^{\underline {abcd} }{\theta ^\delta }{\theta ^\varepsilon }{({\gamma _{\underline {abcd} }})_{\alpha \beta }}{\theta ^\beta } =  - \frac{1}{{3!}}{(\gamma _{}^{\underline {abcd} })_{[\delta \varepsilon }}{({\gamma _{\underline {abcd} }})_{\beta ]\alpha }}{\theta ^\delta }{\theta ^\varepsilon }{\theta ^\beta }\\
T2 = \gamma _{\delta \varepsilon }^{\underline {abc} }{\theta ^\delta }{\theta ^\varepsilon }{({\gamma _{\underline {abc} }})_{\alpha \beta }}{\theta ^\beta } =  - \frac{1}{{3!}}{(\gamma _{}^{\underline {abc} })_{[\delta \varepsilon }}{({\gamma _{\underline {abc} }})_{\beta ]\alpha }}{\theta ^\delta }{\theta ^\varepsilon }{\theta ^\beta }\\
T3 = {C_{\delta \varepsilon }}{\theta ^\delta }{\theta ^\varepsilon }{\theta _\alpha } = \frac{1}{{3!}}{C_{[\delta \varepsilon }}{C_{\beta ]\alpha }}{\theta ^\delta }{\theta ^\varepsilon }{\theta ^\beta }
\end{array}
\end{equation}

In D = 11, any antisymmetric bispinor can be decomposed into a scalar, 3-form, and 4-form. With the basis of ${O^I}_{\varepsilon \beta } = \{ {C_{\varepsilon \beta }},\gamma _{\varepsilon \beta }^{\underline {abc} },\gamma _{\varepsilon \beta }^{\underline {abcd} }\}$, one can expand the above antisymmetric objects and obtain:

\begin{equation}
\begin{array}{l}
{(\gamma _{}^{\underline {abcd} })_{[\delta \varepsilon }}{({\gamma _{\underline {abcd} }})_{\beta ]\alpha }} = \frac{1}{{32}}\{ -7920{C_{[\delta \varepsilon }}{C_{\beta ]\alpha }} - \frac{{528}}{{3!}}{(\gamma _{}^{\underline {abc} })_{[\delta \varepsilon }}{({\gamma _{\underline {abc} }})_{\beta ]\alpha }} + \frac{{144}}{{4!}}{(\gamma _{}^{\underline {abcd} })_{[\delta \varepsilon }}{({\gamma _{\underline {abcd} }})_{\beta ]\alpha }}\} \\
{(\gamma _{}^{\underline {abc} })_{[\delta \varepsilon }}{({\gamma _{\underline {abc} }})_{\beta ]\alpha }} = \frac{1}{{32}}\{ 990{C_{[\delta \varepsilon }}{C_{\beta ]\alpha }} - \frac{{30}}{{3!}}{(\gamma _{}^{\underline {abc} })_{[\delta \varepsilon }}{({\gamma _{\underline {abc} }})_{\beta ]\alpha }} - \frac{{66}}{{4!}}{(\gamma _{}^{\underline {abcd} })_{[\delta \varepsilon }}{({\gamma _{\underline {abcd} }})_{\beta ]\alpha }}\} \\
{C_{[\delta \varepsilon }}{C_{\beta ]\alpha }} = \frac{1}{{32}}\{ {C_{[\delta \varepsilon }}{C_{\beta ]\alpha }} + \frac{1}{{3!}}{(\gamma _{}^{\underline {abc} })_{[\delta \varepsilon }}{({\gamma _{\underline {abc} }})_{\beta ]\alpha }} - \frac{1}{{4!}}{(\gamma _{}^{\underline {abcd} })_{[\delta \varepsilon }}{({\gamma _{\underline {abcd} }})_{\beta ]\alpha }}\} 
\end{array}
\end{equation}

These Fierz identities can be simplified in more compact manner and, consequently, one can see that the terms T1, T2, and T3  are equivalent up to a multiplicative constant.

By checking the other expressions at each level, one can observe such redundancies emerge due to the equivalence of many terms obtained. Moreover, many terms in this expansion route will vanish with the use of Fierz identities, making this approach totally inefficient.

\section{Branching Rules Implementation for Scalar Superfield Decomposition}
In \cite{JamesGates:2019bwu,JamesGates:2020hrl}, the component field content of scalar superfields in D=10 and D=11 was fully extracted based on branching rules.
The splitting of a representation of a Lie algebra $\mathfrak{g}$ into direct sum of irreducible  representations of its Lie subalgebra $\mathfrak{h}$ is called branching rule \cite{Mckay:1977, Slansky:1981yr} and determined by a single projection matrix $P$ in such a way that we have ${v^T}_{\mathfrak{h}} = {P_{}}{._{}}{w^T}_{\mathfrak{g}}$, where ${w}_{\mathfrak{g}}$ is a weight row vector in $\mathfrak{g}$ and ${v}_{\mathfrak{h}}$ is the projected weight vector in $\mathfrak{h}$.  Here, we implement the branching rules in the simplest form, following some parts of the algorithm described in \cite{Fonseca:2020vke}.

To establish a simple branching rules algorithm, one can start with the weights ${W_{R}}$ of a given representation $R = \{ {a_1},{a_2},...,{a_n}\}$ (its Dynkin coefficients) of the Lie algebra $\mathfrak{g}$ (e.g., $\mathfrak{su}(32)$ or $\mathfrak{su}(16)$) which are projected (via projection matrix defined below) into the weights ${V_{R'}}$ of a reducible representation ${R'}$ of the Lie subalgebra $\mathfrak{h}$ (e.g., $\mathfrak{so}(11)$ or $\mathfrak{so}(10)$). Now, from the set of projected weights, one need to find those weights which are the highest weights of irreducible components of the representation ${R'}$.

In order to find the projection matrix for the case $\mathfrak{su}(32) \supset \mathfrak{so}(11)$, we note that there is an obvious branching rule, which is $\{32\}\to\{32\}$.
All of the weights of the representation $\{32\}$ in $\mathfrak{su}(32)$ and $\mathfrak{so}(11)$ can be written in two matrices ${W}_{\mathfrak{su}(32)}$ and ${V}_{\mathfrak{so}(11)}$, where the weights are row vectors. Their transpose is of the form: 

\begin{equation}
\begin{array}{l}
{W^T}_{\mathfrak{su}(32)} = \left( {\left( {\begin{array}{*{20}{c}}
{w{{_1^1}_{\mathfrak{su}(32)}}}\\
 \vdots \\
{w{{_{31}^1}_{\mathfrak{su}(32)}}}
\end{array}} \right)\left( {\begin{array}{*{20}{c}}
{w{{_1^2}_{\mathfrak{su}(32)}}}\\
 \vdots \\
{w{{_{31}^2}_{\mathfrak{su}(32)}}}
\end{array}} \right)...\left( {\begin{array}{*{20}{c}}
{w{{_1^{32}}_{\mathfrak{su}(32)}}}\\
 \vdots \\
{w{{_{31}^{32}}_{\mathfrak{su}(32)}}}
\end{array}} \right)} \right)\\
\\
{V^T}_{\mathfrak{so}(11)} = \left( {\left( {\begin{array}{*{20}{c}}
{v{{_1^1}_{\mathfrak{so}(11)}}}\\
 \vdots \\
{v{{_5^1}_{\mathfrak{so}(11)}}}
\end{array}} \right)\left( {\begin{array}{*{20}{c}}
{v{{_1^2}_{\mathfrak{so}(11)}}}\\
 \vdots \\
{v{{_5^2}_{\mathfrak{so}(11)}}}
\end{array}} \right)...\left( {\begin{array}{*{20}{c}}
{v{{_1^{32}}_{\mathfrak{so}(11)}}}\\
 \vdots \\
{v{{_5^{32}}_{\mathfrak{so}(11)}}}
\end{array}} \right)} \right)\\
\\
\end{array}
\end{equation}

The projection matrix now can be extracted from below:
\begin{equation}
\begin{array}{l}
{V^T}_{\mathfrak{so}(11)} = {P_{}}{._{}}{W^T}_{\mathfrak{su}(32)}
\end{array}
\end{equation}

For computing the weights of a representation of a Lie group, we use the function \textit{Weights} from GroupMath, which is a Mathematica package containing several functions related to Lie Algebras \cite{Fonseca:2020vke}. 
The advanced function \textit{DecomposeRep} in GroupMath, can provide the decomposition based on branching rules. However, for educational purposes and  to be able to access the computation results at each stage of the branching rules, we provide a simple BR-algorithm here.

Let us first look at the weight system of $\{32\}$ in $\mathfrak{su}(32)$
and $\mathfrak{so}(11)$, as follows:
\begin{lstlisting}
Weights[su32, 32]

{{{1, 0, 0, 0, 0, 0, 0, 0, 0, 0, 0, 0, 0, 0, 0, 
0, 0, 0, 0, 0, 0, 0, 0, 0, 0, 0, 0, 0, 0, 0, 0}, 1}, 
{{-1, 1, 0, 0, 0, 0, 0, 0, 0, 0, 0, 0, 0, 0, 0, 
0, 0, 0, 0, 0, 0, 0, 0, 0, 0, 0, 0, 0, 0, 0, 0}, 1},
...
{{0, 0, 0, 0, 0, 0, 0, 0, 0, 0, 0, 0, 0, 0, 0,
0, 0, 0, 0, 0, 0, 0, 0, 0, 0, 0, 0, 0, 0, -1, 1}, 1}, 
{{0, 0, 0, 0, 0, 0, 0, 0, 0, 0, 0, 0, 0, 0, 0,
0, 0, 0, 0, 0, 0, 0, 0, 0, 0, 0, 0, 0, 0, 0, -1}, 1}}

\end{lstlisting}

The function \textit{Weights} returns a list $\{ \{ {w_1},{d_1}\} ,\{ {w_2},{d_2}\} ,...\}$ where the $w$'s are the weights and the $d$'s are their degeneracy. Similarly for the $\mathfrak{so}(11)$, we have:

\begin{lstlisting}
Weights[so11, 32]

{{{0, 0, 0, 0, 1}, 1}, {{0, 0, 0, 1, -1}, 1}, 
{{0, 0, 1, -1, 1}, 1}, {{0, 0, 1, 0, -1}, 1},
...
{{0, 0, -1, 0, 1}, 1}, {{0, 0, -1, 1, -1}, 1}, 
{{0, 0, 0, -1, 1}, 1}, {{0, 0, 0, 0, -1}, 1}}
\end{lstlisting}
We find the explicit projection matrices of $P_{\mathfrak{su}(32) \supset \mathfrak{so}(11)}$
and $P_{\mathfrak{su}(16) \supset \mathfrak{so}(10)}$:

\begin{lstlisting}
<< GroupMath`
W = {};
V = {};
Do[AppendTo[W, Weights[su32, 32][[i, 1]]]; 
AppendTo[V, Weights[so11, 32][[i, 1]]], {i, 32}];
  
WTInv = W.Inverse[Transpose[W].W];
ProjectionMatrix = Transpose[V].WTInv
\end{lstlisting}
\begin{equation}
\hspace*{-1.5cm}
\begin{array}{l}
{P_{\mathfrak{su}(32) \supset \mathfrak{so}(11)}}\\
\\
 = \left( {\scriptsize\begin{array}{*{31}{c}}
0&0&0&0&0&0&0&0&1&2&3&4&5&6&7&8&7&6&5&4&3&2&1&0&0&0&0&0&0&0&0\\
0&0&0&0&1&2&3&4&3&2&1&0&0&0&0&0&0&0&0&0&1&2&3&4&3&2&1&0&0&0&0\\
0&0&1&2&1&0&0&0&0&0&1&2&1&0&0&0&0&0&1&2&1&0&0&0&0&0&1&2&1&0&0\\
0&1&0&0&0&1&0&0&0&1&0&0&0&1&0&0&0&1&0&0&0&1&0&0&0&1&0&0&0&1&0\\
1&0&1&0&1&0&1&0&1&0&1&0&1&0&1&0&1&0&1&0&1&0&1&0&1&0&1&0&1&0&1
\end{array}} \right)\\
\\
\end{array}
\end{equation}

Similarly, for the case $\mathfrak{su}(16) \supset \mathfrak{so}(10)$, we obtain the projection matrix as follows:

\begin{equation}
\hspace*{-1.5cm}
\begin{array}{l}
{P_{\mathfrak{su}(16) \supset \mathfrak{so}(10)}}\\
\\
 = \left( {\scriptsize\begin{array}{*{20}{c}}\small
0&0&0&0&{ - 1}&{ - 2}&{ - 3}&{ - 4}&{ - 3}&{ - 2}&{ - 1}&0&0&0&0\\
0&0&{ - 1}&{ - 2}&{ - 1}&0&0&0&0&0&{ - 1}&{ - 2}&{ - 1}&0&0\\
0&{ - 1}&0&0&0&{ - 1}&0&0&0&{ - 1}&0&0&0&{ - 1}&0\\
{ - 1}&0&0&0&0&0&{ - 1}&0&0&0&{ - 1}&0&{ - 1}&0&0\\
0&0&{ - 1}&0&{ - 1}&0&0&0&{ - 1}&0&0&0&0&0&{ - 1}
\end{array}} \right)
\end{array}
\end{equation}

Now, we look at the component field expansion of a scalar superfield in D=11 at level 7. In order to obtain the expansion at other levels, one can simply replace the parameter \textit{n} with the desired level in the algorithm below.

\begin{lstlisting}
<< GroupMath`
{W, V} = {{}, {}};

Do[AppendTo[W, Weights[su32, 32][[i, 1]]]; 
AppendTo[V, Weights[so11,32][[i, 1]]], {i, 32}];

WTInv = W.Inverse[Transpose[W].W];
ProjectionMatrix = Transpose[V].WTInv;
(* Level-7 *) n = 7 ; 
Representation = SimpleRepInputConversion[su32, 32!/(n! (32 -n)!)] 
W1 = Weights[su32, Representation];
weights = {(ProjectionMatrix.#[[1]]), #[[-1]]} & /@ W1;
weights = DeleteCases[weights, w_ /; UnsameQ[Abs[w[[1]]], w[[1]]]]
\end{lstlisting}

In level 7, we obtain 19,055 positive projected weights, as follows:

\begin{lstlisting}
{{{0,3,0,0,1},1},{{1,1,0,1,1},1},{{1,1,1,0,1},1},
{{1,2,0,0,1},1},{{0,1,0,1,1},1},{{0,1,1,0,1},1},
{{0,2,0,0,1},1}, ...19041... ,{{0,0,0,0,1},1},
{{0,0,0,0,1},1},{{0,0,0,0,1},1},{{0,0,0,0,1},1},
{{0,0,0,0,1},1},{{0,0,0,0,1},1},{{0,0,0,0,1},1}}
\end{lstlisting}

At this stage, we sum over the multiplicities and sort the weights based on their levels, and we get:

\begin{lstlisting}[extendedchars=true,language=Mathematica]
OrderedWeights[CartanMatrix_, weights_] := 
Module[{CmInv, GatherWeights, OrderedWeights},
CmInv = Transpose[Inverse[CartanMatrix]];
GatherWeights = Gather[weights, #1[[1]] === #2[[1]] &];
GatherWeights = Table[{GatherWeights[[i, 1, 1]], 
Total[GatherWeights[[i, All, -1]]]}, {i, Length[GatherWeights]}];
GatherWeights = {#, CmInv.(#[[1]])} & /@ GatherWeights;
OrderedWeights = Sort[GatherWeights, OrderedQ[{#2[[2]], #1[[2]]}] &][[All, 1]]; 
Return[OrderedWeights];]
\end{lstlisting}

\begin{lstlisting}

weights = Flatten[OrderedWeights[so11, #] & /@ {weights}, 1]
\end{lstlisting}

The sorted result is:

\begin{lstlisting}
{{{0,3,0,0,1},1},{{1,1,0,1,1},1},{{1,1,1,0,1},4},{{1,2,0,0,1},13},
{{2,0,0,0,3},3},{{2,0,0,1,1},10},{{2,0,1,0,1},27},{{2,1,0,0,1},64},
{{3,0,0,0,1},138},{{0,0,1,0,3},1},{{0,0,1,1,1},3},{{0,0,2,0,1},9},
{{0,1,0,0,3},10},{{0,1,0,1,1},26},{{0,1,1,0,1},64},{{0,2,0,0,1},148},
{{1,0,0,0,3},62},{{1,0,0,1,1},140},{{1,0,1,0,1},301},{{1,1,0,0,1},615},
{{2,0,0,0,1},1200},{{0,0,0,0,3},296},{{0,0,0,1,1},605},{{0,0,1,0,1},1194},
{{0,1,0,0,1},2277},{{1,0,0,0,1},4216},{{0,0,0,0,1},7627}}
\end{lstlisting}

Now, in order to select the highest weights of irreducible components of the representation ${R'}$ from the set of projected weights, we define two functions, \textit{NonNegativeWeights} and \textit{EliminateWeights}, as:

\begin{lstlisting}
NonNegativeWeights[CartanMatrix_, rep_] := Module[{var1},
var1 = DominantWeights[CartanMatrix, rep[[1]]];
(* weight multiplicity corrections *)
Do[var1[[i]] = {var1[[i, 1]], rep[[-1]]*var1[[i, 2]]},
{i, Length[var1]}];
Return[var1];]
\end{lstlisting}

\begin{lstlisting}
EliminateWeights[TotalSet_, EliminateWeights_] := Module[{var, pos},
var = TotalSet;
Do[pos = Position[TotalSet, EliminateWeights[[i, 1]]][[1, 1]];
var[[pos, 2]] = var[[pos, 2]] - EliminateWeights[[i, 2]];
, {i, Length[EliminateWeights]}];
var = DeleteCases[var, w_ /; w[[2]] == 0];
Return[var];]
\end{lstlisting}

With these functions, one can implement the decomposition as follows:

\begin{lstlisting}
Decomposition = Flatten[Reap[
While[Length[weights] > 0,
Sow[ConstantArray[weights[[1, 1]], weights[[1, -1]]]];
weights = EliminateWeights[weights, NonNegativeWeights[so11,weights[[1]]]];
weights = OrderedWeights[so11, weights]]][[2]], 2]
\end{lstlisting}

The result of the decomposition, in this case, is:

\begin{lstlisting}
{{0,3,0,0,1},{1,1,0,1,1},{1,1,1,0,1},{1,2,0,0,1},{2,0,0,1,1},{2,0,1,0,1},
{2,1,0,0,1},{3,0,0,0,1},{0,0,1,0,3},{0,1,0,0,3},{0,1,0,1,1},{0,1,1,0,1},
{0,1,1,0,1},{0,2,0,0,1},{1,0,0,0,3},{1,0,0,1,1},{1,0,1,0,1},{1,0,1,0,1},
{1,1,0,0,1},{0,0,0,0,3},{0,0,1,0,1},{0,1,0,0,1},{0,0,0,0,1}}
\end{lstlisting}

Here, we use the function \textit{DimR} from GroupMath, which returns the dimension of the representation of the Lie group.

\begin{lstlisting}
Results = Table[DimR[so11, i], {i, Decomposition}]
\end{lstlisting}

\begin{lstlisting}

{264000,1034880,573440,147840,219648,134784,45760,7040,274560,
137280,160160,91520,91520,24960,36960,45056,28512,28512,10240,
4224,3520,1408,32}
\end{lstlisting}

One can see that we have 23 independent component fields at level 7, and that two independent representations $\left\{ 91520,91520\right\}$, each with Dynkin label $\left\{1,0,1,0,1\right\}$, have emerged. This is acceptable only if there are two linearly independent irreducible polynomials at Level 7 as has been argued in \cite{JamesGates:2020hrl}.

In the case of $\mathfrak{su}(m) \supset \mathfrak{so}(n)$, where $\mathfrak{so}(n)$ is the the maximal special subalgebra of $\mathfrak{su}(m)$, there exists a bijection between irreducible representations of $\mathfrak{su}(m)$ and vector partition functions called Plethysms \cite{JamesGates:2020hrl}. In such a situation, the branching rules can be reduced to a more efficient algorithm for the decomposition.

\section{Bosonic and Fermionic Superfields Decomposition}
With the efficient branching rules method for superfield component decomposition, one can now move forward and search for the traceless graviton and the traceless gravitino not just within scalar superfields, but by scanning of the full spectrum of the component fields of bosonic and fermionic superfields \cite{JamesGates:2019bwu,JamesGates:2020hrl}.

Consider $[indices]$ as any combination of bosonic and fermionic indices. One can attach these indices to the scalar superfield and obtain:

\begin{equation}
\begin{array}{l}
{S_{[indices]}}({x^{\underline k}},{\theta ^\alpha }) = {s^{(0)}}_{[indices]}({x^{\underline k}}) + {\theta ^\alpha }{s_{\alpha [indices]}}^{(1)}({x^{\underline k}}) + {\theta ^\alpha }{\theta ^\beta }{s_{\alpha \beta [indices]}}^{(2)}({x^{\underline k}}) +  \cdot  \cdot  \cdot 
\end{array}
\end{equation}

For simplicity, suppose $[indices]$ corresponds to only one irreducible representation, denoted by $\{ {\mathop{\rm Re}\nolimits} p\}$. If we denote level n component field content of scalar superfield by ${C_n}$, then the level n component field content of $ S_{[indices]}({x^{\underline k}},{\theta ^\alpha })$ will be ${C_n} \otimes \{ {\mathop{\rm Re}\nolimits} p\}$. With this in mind the scalar superfield $ S({x^{\underline k}},{\theta ^\alpha })$ can be seen as a means to obtain the explicit component spectrum of all possible superfields as has been mentioned in \cite{JamesGates:2020hrl}.
This technology can facilitate the search for possible off-shell supergravity supermultiplets. For example, the irreducible representations of the traceless graviton and traceless gravitinos in D=10, $\mathcal{N}=1$ are $\{54\}$ and $\{144\}$, respectively. As studied in \cite{JamesGates:2019bwu}, the bosonic superfield $S({x^{\underline k}},{\theta ^\alpha })\otimes\{ 120\}$ provides 4 possibilities for the embedding of traceless gravitons at levels $2,6, 10$, and $14$, . If a tracless gravition emerges at the level n, then we have a traceless gravitino at level (n+1).
Here, we re-derive the levels 6 and 7 component field of $S({x^{\underline k}},{\theta ^\alpha }) \otimes \{ 120\}$ in which $\{ 120\}$ is a bosonic irreducible representation.
In the branching rules algorithm provided in the previous section, we insert $\mathfrak{su}(16)$ as our Lie algebra and $\mathfrak{so}(10)$ as the Lie subalgebra. Furthermore, we set the level parameter to n = 6 for the level 6 and n=7 for the level 7 and insert the correct level formula $\frac{{16!}}{{n!\left( {16 - n} \right)!}}$.

At level 6, the non-negative projected weights of the scalar superfield are:
\begin{lstlisting}
{{{2, 0, 1, 0, 0}, 1}, {{3, 0, 0, 0, 0}, 4}, {{0, 1, 0, 2, 0}, 1}, 
{{0, 1, 1, 0, 0}, 2}, {{1, 0, 0, 1, 1}, 6}, {{1, 1, 0, 0, 0}, 16}, 
{{0, 0, 0, 2, 0}, 16}, {{0, 0, 0, 0, 2}, 12}, {{0, 0, 1, 0, 0},34}, 
{{1, 0, 0, 0, 0}, 72}}
\end{lstlisting}

We obtain the decomposition results, in terms of Dynkin labels, as follows:

\begin{lstlisting}
{{2, 0, 1, 0, 0}, {0, 1, 0, 2, 0}}
\end{lstlisting}
The dimensions of the irreducible representations are:
\begin{lstlisting}
{4312, 3696}
\end{lstlisting}

Next, we use LieART \cite{Feger:2019tvk} for the tensor product decomposition of $\{ 4312\}  \otimes \{ 120\}$ and $\{ 3696\}  \otimes \{ 120\}$ as follows:

\begin{lstlisting}
<< LieART`
DecomposeProduct[Irrep[SO10][4312], Irrep[SO10][120]]
\end{lstlisting}

\begin{equation}
\begin{array}{l}
\{54\}+\{660\}+\{770\}+\{945\}+\{1050\}+\overline{\{1050\}}+2 \{1386\}+\{4125\}+\{5940\}\\+3 \{8085\}
+\{14784\}+\{16380\}+\{17325\}+\overline{\{17325\}}+2 \{17920\}+\{23040\}+\overline{\{23040\}}\\+\{72765\}
+\{112320\}+\{143000\}
\end{array}
\end{equation}

\begin{lstlisting}
DecomposeProduct[Irrep[SO10][IrrepPrime[3696]], Irrep[SO10][120]]
\end{lstlisting}

\begin{equation}
\begin{array}{l}
\{210\}+\{770\}+\{945\}+2 \{1050\}+\{2772\}+\{4125\}+2 \{5940\}+2 \{6930\}\\
+\{8085\}+\{8910\}+\{17920\}+2 \{23040\}+\{50688\}+\{72765\}+\{73710\}+\{128700\}
\end{array}
\end{equation}

By adding the results in (10) and (11), we obtain level 6 component decomposition for our bosonic superfield. As shown, the traceless graviton $\{54\}$ emerged at level 6.

Similarly, for level 7, we have the non-negative projected weights of the scalar superfield as:
\begin{lstlisting}
{{{3, 0, 0, 0, 1}, 1}, {{1, 0, 1, 1, 0}, 1}, {{1, 1, 0, 0, 1}, 3},
{{2, 0, 0, 1, 0}, 10}, {{0, 0, 0, 2, 1}, 3}, {{0, 0, 1, 0, 1}, 7},
{{0, 1, 0, 1, 0}, 19}, {{1, 0, 0, 0, 1}, 42}, {{0, 0, 0, 1, 0}, 85}}
\end{lstlisting}

The decomposition result is:
\begin{lstlisting}
{{3, 0, 0, 0, 1}, {1, 0, 1, 1, 0}}
\end{lstlisting}

\begin{lstlisting}
{2640, 8800}
\end{lstlisting}

Next, in order to extract component field of $S({x^{\underline k}},{\theta ^\alpha }) \otimes \{ 120\}$, we perform the below tensor product decompositions: 

\begin{lstlisting}
DecomposeProduct[Irrep[SO10][120], Irrep[SO10][Bar[8800]]]
\end{lstlisting}

\begin{equation}
\begin{array}{l}
\{144\}+\{560\}+\{672\}+\{720\}+2 \{1200\}+\{1440\}+\{2640\}+3 \{3696\}
+\{8064\}\\
+3 \{8800\}+3 \{11088\}+\{15120\}+\{17280\}+\{23760\}+3 \{25200\}+\{30800\}
+\{34992\}\\
+2 \{38016\}+\{43680\}+\{49280\}+\{55440\}+\{144144\}+\{196560\}+\{205920\}
\end{array}
\end{equation}

\begin{lstlisting}
DecomposeProduct[Irrep[SO10][120], Irrep[SO10][Bar[2640]]]
\end{lstlisting}

\begin{equation}
\begin{array}{l}
\{720\}+2 \{2640\}+\{3696\}+\{7920\}+\{8800\}+2 \{15120\}\\+\{38016\}+\{48048\}+\{49280\}+\{124800\}
\end{array}
\end{equation}
By adding the results in (12) and (13), one can obtain the level 7 component decomposition for the bosonic superfield. It can be seen that the traceless gravitino $\{144\}$ emerged at level 7.

Similarly, one can investigate fermionic superfields by considering $[indices]$ to be spinor indices. For instance, consider $S({x^{\underline k}},{\theta ^\alpha }) \otimes \{ 144\}$, in which $\{ 144\}$ is an spinorial irrep. The component fields of this fermionic superfield at level 3 that contain the traceless graviton is:

\begin{lstlisting}
DecomposeProduct[Irrep[SO10][144], Irrep[SO10][Bar[560]]]
\end{lstlisting}
\begin{equation}
\begin{array}{l}
\{45\}+\{54\}+2\{210\}+\{770\}+3\{945\}+2\{1050\}+\overline{\{1050\}}+\{1386\}+\{4125\}\\+2\{5940\}+\{6930\}+\{8085\}+\{17920\}+\{23040\}
\end{array}
\end{equation}

\section*{Acknowledgments}
The author thanks S. James Gates, Jr., Y. Hu and S.-N. Mak, for their helpful discussion and comments.

\section*{Appendix: Useful Identities for 11D Gamma Matrices}
We used the following identities and Fierz general formula in expanding antisymmetric expressions into basis.  
\begin{equation}
\begin{array}{l}
{\gamma ^{\underline {ijkl} }}{\gamma ^{\underline {mnop} }}{\gamma _{\underline {ijkl} }} =  - 144{\gamma ^{\underline {mnop} }}\\
{\gamma ^{\underline {ijkl} }}{\gamma ^{\underline {mno} }}{\gamma _{\underline {ijkl} }} =  - 528{\gamma ^{\underline {mno} }}\\
{\gamma ^{\underline {ijk} }}{\gamma ^{\underline {lmn} }}{\gamma _{\underline {ijk} }} =  - 30{\gamma ^{\underline {lmn} }}\\
{\gamma ^{\underline {ijk} }}{\gamma ^{\underline {lmno} }}{\gamma _{\underline {ijk} }} = 66{\gamma ^{\underline {lmno} }}\\
{\gamma ^{\underline {ijkl} }}{\gamma _{\underline {ijkl} }} = 7920\\
{\gamma ^{\underline {ijk} }}{\gamma _{\underline {ijk} }} =  - 990
\end{array}
\end{equation}

\begin{equation}
{M_{\delta \varepsilon }}{N_{\beta \alpha }} = {2^{ - [D/2]}}\sum\limits_I^{} {{{({M_{}}{{\rm O}_I}N)}_{\delta \alpha }}{O^I}_{\varepsilon \beta }} 
\end{equation}

Where the basis ${O^I}_{\varepsilon \beta }$ in our case are ${C_{\varepsilon \beta }},\gamma _{\varepsilon \beta }^{\underline {abc} }$ and $\gamma _{\varepsilon \beta }^{\underline {abcd}}$.

\end{document}